\documentclass{elsart}
\begin{document}
\begin{flushright}
KUPT 02-02 \\
May 2002
\end{flushright}
\begin{frontmatter}
\title{Symmetries in collinear effective theory}
\author{Junegone Chay\thanksref{email}}, 
\thanks[email]{e-mail address: chay@korea.ac.kr}
\author{Chul Kim}
\address{Department of Physics, Korea University, Seoul 136-701,
Korea} 
\begin{abstract}
We consider various symmetries present in the collinear effective
theory and their implications. There are collinear, soft and ultrasoft
gauge symmetries and we discuss transformation properties of a
collinear quark and gauge fields under these gauge transformations. 
If we require the gauge invariance order by order, the gauge fields
have definite transformation properties. We also probe
reparameterization invariance and residual energy invariance in the
collinear effective theory and discuss their physical consequences.  

\end{abstract}
\end{frontmatter}

When we describe strong interaction processes which include massless
particles with large energy $E$, we can use the collinear effective
theory \cite{bauer1,bauer2,bauer3,bauer4}. In the 
collinear effective theory, the effective Lagrangian can be
systematically expanded in powers of $\lambda \sim p_{\perp}/E$ where
$p_{\perp}$ is the transverse momentum of a collinear quark. Even
though there is one large parameter $E$, there are three scales
involved in the effective theory. The
momentum of an energetic particle can be written as
$p^{\mu} = \half(\overline{n} \cdot p) n^{\mu} + p_{\perp}^{\mu} +
 \half (n \cdot p) \overline{n}^{\mu}$,   
where $\overline{n} \cdot p\sim \lambda^0$, $p_{\perp} \sim \lambda$,
and $n\cdot p \sim \lambda^2$. If we include small fluctuations due to
the strong interaction, the momentum $P^{\mu}$ of an energetic
particle is decomposed as the sum of a large momentum $p^{\mu}$ with
$\overline{n}\cdot p \sim \lambda^0$, $p_{\perp}^{\mu} \sim \lambda$
and $n\cdot p=0$ and a small residual momentum $k^{\mu}\sim\lambda^2$:
\begin{equation}
  \label{eq:momentum}
P^{\mu} = p^{\mu} + k^{\mu} = \frac{n^{\mu}}{2} \overline{n}\cdot p +
p_{\perp}^{\mu} +k^{\mu}.  
\end{equation}
Here $p^{\mu}$ acts as a label momentum on collinear fields, and the
residual momentum $k^{\mu}$ is a small fluctuation. 

Since there are three distinct scales $E$, $E\lambda$, $E\lambda^2$, 
we have to construct
effective theories at each scale. Here we consider an effective theory
in the region $E\lambda < \mu < E$, in which collinear degrees of
freedom above $\mu$ are integrated out. This effective theory can be
used in exclusive 
decays of a heavy meson into energetic light mesons. It has been
applied to the resummation of Sudakov logarithms in $B$ decays
\cite{bauer1} and to the proof of factorization in
$\overline{B}\rightarrow D\pi$ \cite{bauer5}. And  
the calculation of the form factors in heavy-to-light transitions to
order $\lambda$ was performed \cite{chay1}.   

The symmetry structure in the collinear effective theory is
rich. There are interesting physical implications by combining the
symmetries of the effective theory. In this Letter, we consider
gauge invariance, reparameterization invariance and residual energy
invariance.

In order to discuss gauge symmetries of the collinear effective
theory, let us consider the full QCD first. The Lagrangian for a
massless quark and gluons is given by
\begin{equation}
\mathcal{L}_{\mathrm{QCD}} = \overline{q} i\FMSlash{D} q
-\frac{1}{4} G_{\mu\nu}^a G^{\mu\nu a},
\end{equation}
where the covariant derivative $D^{\mu}$ is defined as $D_{\mu} =
\partial_{\mu} +igA_{\mu}^a T^a$ and $G_{\mu\nu}^a$ is the gluon field
strength tensor. The quark field and the gauge fields transform under
an arbitrary $SU(3)$ gauge transformation of
the form $U=e^{i\omega^a (x) T^a}$ as
\begin{equation}
  \label{eq:quark}
q(x)\rightarrow U(x)q(x), \ \ A^{\mu}(x) \rightarrow U
(x)A^{\mu} (x) U^{\dagger}(x)
-\frac{i}{g} U (x) \partial^{\mu} U^{\dagger} (x),
\end{equation}
and the Lagrangian is invariant under the gauge transformation.

In the collinear effective theory, we decompose each field into
a collinear field, a soft field and an ultrasoft (usoft) field. For a
collinear quark, we redefine the field by extracting large momenta as
\begin{equation}
  \label{eq:coquark}
q_{n,p} (x) = \sum_{p} e^{-ip\cdot x} q(x),
\end{equation}
where $p^{\mu} = \half (\overline{n}\cdot p)  n^{\mu} + p_{\perp}^{\mu}$
is a large label momentum to order $\lambda$. And using a projection
operator $\FMslash{n}\overline{\FMslash{n}} /4$ to extract a large
component in the $n^{\mu}$ direction, we can express the effective
Lagrangian for collinear quarks in terms of $\xi_n$, which is defined
as
\begin{equation}
\xi_n = \frac{\FMslash{n} \overline{\FMslash{n}} }{4} q_{n,p} \
\mbox{with}\ \FMslash{n} \xi_n =0. 
\end{equation}
We omit the subscripts for the label momentum in $\xi_n$ for
simplicity. Since we extract the large momentum, the derivative acting
on $\xi_n$ produces momentum of order $\lambda^2$. In order to
facilitate the power counting in $\lambda$, it is convenient to
define label operators $\overline{\mathcal{P}}$ and
$\mathcal{P}_{\perp}^{\mu}$, which 
pick out the label momenta of collinear fields of order $\lambda^0$
and $\lambda$ respectively. For example, $\overline{\mathcal{P}}
\xi_n = \overline{n} \cdot p\, \xi_n$. Then the operator which
picks out label momenta can be written as $\mathcal{P}^{\mu} =
\half \overline{\mathcal{P}} n^{\mu} + \mathcal{P}_{\perp}^{\mu}$. 
The gauge field $A_{\mu}$ is also decomposed as
$A^{\mu} = A_c^{\mu} + A_s^{\mu} +A_{u}^{\mu}$,
where $A_c^{\mu}$ is a collinear gauge field, $A_s^{\mu}$ is a soft
gauge field and $A_{u}^{\mu}$ is an usoft gauge
field.

The effective Lagrangian of the collinear quark sector in the
collinear effective theory is given by 
\begin{eqnarray}
  \label{eq:eff}
\mathcal{L} &=&  \overline{\xi}_n \Bigl\{ n\cdot
(iD -g A_n)  \nonumber \\
&&+(\FMslash{\mathcal{P}}_{\perp}
-g\FMslash{A}_n^{\perp} +i\FMslash{D}_{\perp} )
\frac{1}{\overline{n}\cdot (\mathcal{P}-gA_n +iD)} 
(\FMslash{\mathcal{P}}_{\perp}-g\FMslash{A}_n^{\perp}+i\FMslash{D}_{\perp}) 
\Bigr\} \frac{\FMslash{\overline{n}}}{2} \xi_n,
\end{eqnarray}
where we write the collinear gauge field $A_c^{\mu} = \sum_q
e^{-iq\cdot x} A_n^{\mu}$ explicitly, and the covariant derivative
contains only usoft gluons. The summation on the label momenta and the
phases are omitted. Note that there appear no soft gluons in the
collinear quark sector because the interaction of a soft gluon with a
collinear quark makes the collinear quark off the mass shell.

From all the possible gauge transformation in the full QCD, the 
relevant ones in constructing the effective theory are those which
have support over collinear, soft, and ultrasoft momenta. A
collinear gauge transformation $U_c (x)$ is defined as the subset of
gauge transformations where $\partial_{\mu} U_c (x) \sim
(\lambda^2,1,\lambda)$. A soft gauge transformation $U_s$ is the
subset of gauge transformations with $\partial_{\mu} U_s (x) \sim
\lambda$, and an usoft gauge transformation $U_u$ is the subset with
$\partial_{\mu} U_u \sim \lambda^2$. And the collinear, soft, and
usoft gluons are the gauge fields associated with these
transformations.

Now we specify the transformation properties of all the fields in the
effective theory under these gauge transformations We require
that the effective Lagrangian is invariant under these gauge
transformations. Here the point is that we can assign 
arbitrary transformation properties to all the fields as long as the
sum of the gauge fields $A_c^{\mu} +A_s^{\mu} + A_u^{\mu}$ transforms
as
\begin{equation}
  \label{eq:sumgauge}
A_c^{\mu} +A_s^{\mu} + A_u^{\mu} \rightarrow U_i (A_c^{\mu} +A_s^{\mu}
+ A_u^{\mu}) U_i^{\dagger} -\frac{i}{g} U_i \partial^{\mu}
U_i^{\dagger}, \ \ (i=c,s,u).
\end{equation}
This is because the gauge transformations $U_i$ are subsets of the
gauge transformations of the full QCD and $A_c^{\mu} +A_s^{\mu} +
A_u^{\mu}$ is the gauge field in the full QCD. Therefore there is
arbitrariness in specifying the gauge transformation properties of
each gauge field. However, if we require that the effective Lagrangian
be invariant at each order in $\lambda$, the gauge transformation for
each field becomes more specific. 

Let us consider one possible choice in which the gauge fields
transform under collinear gauge transformations as
\begin{equation}
  \label{eq:co1}
A_c^{\mu} \rightarrow U_c A_c^{\mu} U_c^{\dagger} -\frac{i}{g} U_c
\partial^{\mu} U_c^{\dagger}, \ \
A_s^{\mu} \rightarrow U_c A_s^{\mu} U_c^{\dagger}, \ \
A_u^{\mu} \rightarrow U_c A_u^{\mu}  U_c^{\dagger}.
\end{equation}
Here we put the inhomogeneous term coming from the derivative of the
gauge transformation in the transformation of a collinear gauge
field. The full effective Lagrangian in Eq.~(\ref{eq:eff}) is
invariant under the gauge transformation in Eq.~(\ref{eq:co1}), but it
does not make the Lagrangian gauge invariant at each order in
$\lambda$. In order to show this fact in a transparent way, let us
write the collinear gauge transformation $U_c$ as
\begin{equation}
  \label{eq:coltr}
U_c = \sum_Q e^{-iQ\cdot x} \mathcal{U},
\end{equation}
where $Q^{\mu}$ is the label momentum and $\partial^{\mu}
\mathcal{U}\sim \lambda^2$. Then the gauge transformation in
Eq.~(\ref{eq:co1}) becomes 
\begin{equation}
  \label{eq:co2}
A_n^{\mu} \rightarrow \mathcal{U} A_n^{\mu} \mathcal{U}^{\dagger}
-\frac{1}{g} \mathcal{U} \Bigl[ (\mathcal{P}^{\mu} +
i\partial^{\mu} ) \mathcal{U}^{\dagger} \Bigr], \ \
A_s^{\mu} \rightarrow \mathcal{U} A_s^{\mu} \mathcal{U}^{\dagger}, \ \  
A_u^{\mu} \rightarrow \mathcal{U} A_u^{\mu} \mathcal{U}^{\dagger},
\end{equation}
where the square bracket means that the operator
acts only inside the bracket. And the collinear spinor $\xi_n$
transforms as $\xi_n \rightarrow \mathcal{U} \xi_n$. 
Under this gauge transformation, $\mathcal{P}^{\mu} -gA_n^{\mu} +iD^{\mu}$
transforms as
\begin{equation}
  \label{eq:combcol}
\mathcal{P}^{\mu} -gA_n^{\mu} +iD^{\mu} \rightarrow \mathcal{U} \Bigl(
\mathcal{P}^{\mu}   
-gA_n^{\mu} +iD^{\mu} \Bigr) \mathcal{U}^{\dagger}.  
\end{equation}
Note that the full Lagrangian contains only the combination
$\mathcal{P}^{\mu} -gA_n^{\mu} +iD^{\mu}$ contracted with $n^{\mu}$,
$\overline{n}^{\mu}$ or $\gamma_{\perp}^{\mu}$. Therefore the whole
Lagrangian is invariant 
under the gauge transformation of Eq.~(\ref{eq:co2}). However this
gauge transformation does not make the Lagrangian invariant order by
order in $\lambda$. 

We can expand the effective Lagrangian in Eq.~(\ref{eq:eff}) in powers
of $\lambda$ as $\mathcal{L} =\mathcal{L}_0 + \mathcal{L}_1
+\mathcal{L}_2 + \cdots$, where
\begin{eqnarray}
  \label{eq:lamlag}
\mathcal{L}_0 &=&  \overline{\xi}_n \Bigl\{ n\cdot
(iD +\mathcal{P}-g A_n) + \Bigl(\FMslash{\mathcal{P}}_{\perp}
-g\FMslash{A}_n^{\perp} \Bigr)  
\frac{1}{\overline{n} \cdot ({\mathcal{P}-gA_n})} 
\Bigl(\FMslash{\mathcal{P}}_{\perp}-g\FMslash{A}_n^{\perp} \Bigr)
\Bigr\}  
\frac{\FMslash{\overline{n}}}{2} \xi_n, \nonumber \\
\mathcal{L}_1 &=&  \overline{\xi}_n \Bigl\{ i\FMslash{D}_{\perp} 
\frac{1}{\overline{n} \cdot ({\mathcal{P}}-gA_n)} 
\Bigl(\FMslash{\mathcal{P}}_{\perp}-g\FMslash{A}_n^{\perp} \Bigr) 
+\Bigl(\FMslash{\mathcal{P}}_{\perp} -g\FMslash{A}_n^{\perp} \Bigr)  
\frac{1}{\overline{n} \cdot ({\mathcal{P}}-gA_n)}
i\FMslash{D}_{\perp} 
\Bigr\} \frac{\FMslash{\overline{n}}}{2} \xi_n, \nonumber \\
\mathcal{L}_2 &=&  \overline{\xi}_n \Bigl\{ i\FMslash{D}_{\perp} 
\frac{1}{\overline{n} \cdot ({\mathcal{P}}-gA_n)} i\FMslash{D}_{\perp}
\nonumber \\
&& -\Bigl(\FMslash{\mathcal{P}}_{\perp}-g\FMslash{A}_n^{\perp} \Bigr)
\frac{1}{\overline{n} \cdot (\mathcal{P} -gA_n)} \overline{n}
\cdot iD \frac{1}{\overline{n} \cdot (\mathcal{P}-gA_n)}  
\Bigl(\FMslash{\mathcal{P}}_{\perp}-g\FMslash{A}_n^{\perp}
\Bigr) \Bigr\} \frac{\FMslash{\overline{n}}}{2} \xi_n. 
\end{eqnarray}
From this expansion, it is clear why the gauge transformation of
Eq.~(\ref{eq:co2}) does not preserve the gauge invariance order by
order. At each order, the operators appear in the combinations
$n\cdot (\mathcal{P} -gA_n +iD)$, $\FMSlash{\mathcal{P}}_{\perp} - 
g\FMSlash{A}_n^{\perp}$, $i\FMSlash{D}_{\perp}$, $\overline{n}\cdot
(\mathcal{P}-gA_n)$ and $\overline{n}\cdot iD$. Therefore the basic
combinations of the operators which appear
in the Lagrangian are $\mathcal{P}^{\mu} -gA_n^{\mu}$ and $iD^{\mu}$,
contracted with $n^{\mu}$, $\overline{n}^{\mu}$ or
$\gamma_{\perp}^{\mu}$, and they transform as
\begin{eqnarray}
  \label{eq:ordgau}
\mathcal{P}^{\mu} -gA_n^{\mu} &\rightarrow& \mathcal{U}
(\mathcal{P}^{\mu} -gA_n^{\mu} )\mathcal{U}^{\dagger} +\mathcal{U}
\Bigl[i\partial^{\mu} \mathcal{U}^{\dagger}\Bigr], 
\nonumber  \\
iD^{\mu} &\rightarrow&  \mathcal{U}iD^{\mu}  \mathcal{U}^{\dagger} -
\mathcal{U} \Bigl[i\partial^{\mu} \mathcal{U}^{\dagger} \Bigr].  
\end{eqnarray}
The combination $\mathcal{P}^{\mu} -gA_n^{\mu}+iD^{\mu}$ transforms
homogeneously, but the decompositions $\mathcal{P}^{\mu} -gA_n^{\mu}$
and $iD^{\mu}$ have additional 
terms, which are opposite in sign. Therefore though the whole
Lagrangian is invariant under this gauge transformation, the
Lagrangian at each order is not invariant
under the transformation of Eq.~(\ref{eq:co2}). 

As a special example, we can choose $\mathcal{U}$ such that
$\partial^{\mu} \mathcal{U}$ has only the component in the
$\overline{n}^{\mu}$ direction. That is, gluons transform
under a collinear transformation as
\begin{eqnarray}
\label{eq:bauer}
A_n^{\mu} &\rightarrow& \mathcal{U} A_n^{\mu} \mathcal{U}^{\dagger}
-\frac{1}{g} \mathcal{U} \Bigl[ (\mathcal{P}^{\mu}
+\frac{\overline{n}^{\mu}}{2} in\cdot \partial) \mathcal{U} \Bigr],
\nonumber \\ 
A_s^{\mu} &\rightarrow& \mathcal{U} A_s^{\mu} \mathcal{U}^{\dagger}, \
A_u^{\mu} \rightarrow \mathcal{U} A_u^{\mu} \mathcal{U}^{\dagger}. 
\end{eqnarray}
In this case, since $\partial^{\mu} \mathcal{U}$ has only the
component in the $\overline{n}^{\mu}$ direction, the effective
Lagrangian is collinear gauge invariant order by order \cite{chay1}. 
Though the collinear gauge transformation in Eq.~(\ref{eq:bauer}) is a
legitimate gauge transformation, it is a very restricted subset of the
most general transformations with $\partial^{\mu} \mathcal{U} \sim
\lambda^2$. For the most general transformations of $\mathcal{U}$, we
can write the derivative of $\mathcal{U}$ as
\begin{equation}
  \label{eq:general}
\partial^{\mu} \mathcal{U} = \frac{n^{\mu}}{2} \overline{n}\cdot
\partial \mathcal{U} + \partial_{\perp}^{\mu} \mathcal{U} +
\frac{\overline{n}^{\mu}}{2} n\cdot \partial \mathcal{U},  
\end{equation}
where each term is of order $\lambda^2$. 

If we use this most general collinear gauge transformation in
Eq.~(\ref{eq:co2}), the collinear gauge transformation 
is troublesome if we describe gauge-invariant operators 
such as heavy-to-light current operators at each order in
$\lambda$. In general, we cannot make gauge-invariant operators at
fixed order in $\lambda$. In order to make an operator gauge
invariant, we have to invoke operators with different powers of
$\lambda$.  

In Ref.~\cite{bauer4}, they have considered the gauge symmetry of the
leading-order Lagrangian. They also argue for physical reasons
that the usoft field acts as a background field and they extend the
transformation of gluons in the presence of a background usoft
gluon. Following the background field method of Abbott \cite{abbott},
the correct transformation property in the background usoft 
field, with the assumption that $\partial^{\mu}\mathcal{U}$ has only
the $\overline{n}^{\mu}$ component, is given by
\begin{eqnarray}
  \label{eq:bauer1}
A_n^{\mu} &\rightarrow& \mathcal{U} A_n^{\mu} \mathcal{U}^{\dagger}
-\frac{1}{g}  \mathcal{U}
[(\mathcal{P}^{\mu}+\frac{\overline{n}^{\mu}}{2} n\cdot iD)
\mathcal{U}^{\dagger}] -\frac{\overline{n}^{\mu}}{2} n\cdot A_u,
\nonumber  \\  
A_s^{\mu}&\rightarrow& A_s^{\mu}, \ \ A_u^{\mu} \rightarrow A_u^{\mu}, 
\end{eqnarray}
replacing the derivative with a covariant derivative including a usoft
background gluon. Under the collinear gauge transformation in
Eq.~(\ref{eq:bauer1}), only the leading Lagrangian $\mathcal{L}_0$ is
invariant, but the Lagrangians $\mathcal{L}_1$, $\mathcal{L}_2$ are not
invariant. Therefore the full Lagrangian is not gauge invariant. This
is because the combination $\mathcal{P}^{\mu} -gA_n^{\mu} + iD^{\mu}$
does not transform homogeneously under the transformation in
Eq.~(\ref{eq:bauer1}), but only the combinations $n\cdot (\mathcal{P}
-gA_n +iD)$ and $\FMSlash{\mathcal{P}}_{\perp} -g\FMSlash{A}_{\perp}$,
which appear in $\mathcal{L}_0$, transform homogeneously. However, those
terms such as $i\FMSlash{D}_{\perp}$ and $\overline{n}\cdot iD$, which
appear in the Lagrangian at higher order, do not transform
homogeneously.

If we want to keep the Lagrangian gauge invariant at each order in
$\lambda$, we require the gauge fields transform under a collinear
gauge transformation as
\begin{eqnarray}
  \label{eq:truegauge}
A_n^{\mu} &\rightarrow& \mathcal{U} A_n^{\mu} \mathcal{U}^{\dagger}
-\frac{1}{g} \mathcal{U} \Bigl[\mathcal{P}^{\mu}
\mathcal{U}^{\dagger}\Bigr],  \ \
A_s^{\mu} \rightarrow \mathcal{U} A_s^{\mu} \mathcal{U}^{\dagger},
\nonumber  \\ 
A_u^{\mu} &\rightarrow& \mathcal{U} A_u^{\mu} \mathcal{U}^{\dagger}
-\frac{i}{g} \mathcal{U} \Bigl[\partial^{\mu} \mathcal{U}^{\dagger}
\Bigr].   
\end{eqnarray}
One may wonder why usoft fields transform inhomogeneously under
collinear gauge transformations since they fluctuate over wavelengths
which cannot resolve the fast local change induced by $U_c(x)$. But
this is  true only when $\partial^{\mu} \mathcal{U}=0$, that is, when
the collinear gauge transformation does not include any fluctuation of
order $\lambda^2$ ($\partial^{\mu} \mathcal{U} \sim \lambda^2$). If we
include small fluctuations of order $\lambda^2$ which is in
superposition with the fast local change, usoft gluons can
also change under this small remnant fluctuation of the collinear
gauge transformation. Note that the gauge transformation
shown in Eq.~(\ref{eq:truegauge}) is not unique. We can devise other
gauge transformations which make the effective Lagrangian invariant
order by order. Eq.~(\ref{eq:truegauge}) is one possible choice which
satisfies this requirement.

Under the gauge transformations given by Eq.~(\ref{eq:truegauge}),
the decompositions $\mathcal{P}^{\mu} -gA_n^{\mu}$ and $iD^{\mu}$ 
transform homogeneously as 
\begin{equation}
  \label{eq:corgau}
\mathcal{P}^{\mu} -gA_n^{\mu} \rightarrow \mathcal{U}
(\mathcal{P}^{\mu} -gA_n^{\mu} )\mathcal{U}^{\dagger}, \ \ 
iD^{\mu} \rightarrow \mathcal{U}iD^{\mu}  \mathcal{U}^{\dagger}.    
\end{equation}
Since the effective Lagrangian contains either $\mathcal{P}^{\mu}
-gA_n^{\mu}$ or $iD^{\mu}$ at each order, the effective Lagrangian is
invariant order by order under the collinear 
gauge transformation given by Eq.~(\ref{eq:truegauge}). Then it is
possible to consider gauge-invariant operators at any fixed order in
$\lambda$.

We can choose
a different gauge such as the background gauge. It is convenient
in calculating radiative corrections for the operators with
external gluons. If we are interested in the operators with external
usoft gluons,  we can write the usoft field as $A_u^{\mu} = B_u^{\mu}
+ Q_u^{\mu}$, 
where $B_u^{\mu}$ is the background usoft field and $Q_u^{\mu}$ is the
quantum usoft field. If we also require the collinear
gauge invariance order by order in $\lambda$, each gauge field
transforms as
\begin{eqnarray}
  \label{eq:backgauge}
A_n^{\mu} &\rightarrow& \mathcal{U} A_n^{\mu} \mathcal{U}^{\dagger}
-\frac{1}{g} \mathcal{U} \Bigl[\mathcal{P}^{\mu}
\mathcal{U}^{\dagger}\Bigr], \ \
A_s^{\mu} \rightarrow \mathcal{U} A_s \mathcal{U}^{\dagger}, \nonumber 
\\ 
Q_u^{\mu} &\rightarrow& \mathcal{U} Q_u^{\mu} \mathcal{U}^{\dagger}
-\frac{i}{g} \mathcal{U} \Bigl[\tilde{D}^{\mu}
\mathcal{U}^{\dagger}]-B_u^{\mu},  
\end{eqnarray}
where $\tilde{D}^{\mu} = \partial^{\mu} + igB_u^{\mu}$
\cite{abbott}. Note that we cannot neglect the quantum degrees of
freedom for usoft fields as long as we include small fluctuations of
order $\lambda^2$ in the collinear gauge transformation.

Since the soft gauge transformation is not relevant in the collinear
quark sector, we will consider the remaining usoft gauge
transformation. Under a usoft gauge transformation, all the fields
transform as in QCD. Gauge fields transform under usoft gauge
transformations as
\begin{equation}
  \label{eq:usoft}
 A_n^{\mu} \rightarrow U_u A_n^{\mu} U_u^{\dagger},  \ \
A_s^{\mu} \rightarrow  U_u A_s^{\mu} U_u^{\dagger}, \ \
A_u^{\mu} \rightarrow  U_u A_u^{\mu} U_u^{\dagger} -\frac{i}{g} U_u
\partial^{\mu} U_u^{\dagger},
\end{equation}
and the spinor $\xi_n$ transforms as $\xi_n \rightarrow U_u \xi_n$. 
The effective Lagrangian is invariant under the usoft gauge
transformation in Eq.~(\ref{eq:usoft}) order by order. 

Next let us consider reparameterization invariance of the collinear
effective theory. Reparameterization invariance appears whenever we
decompose momenta into large and small components. For instance,
in the heavy quark effective theory (HQET), we can decompose the
momentum of a heavy quark as $p_Q^{\mu} = m_Q v^{\mu} + k^{\mu}$,
where $k^{\mu}$ is the residual momentum of order
$\Lambda_{\mathrm{QCD}}$. We can shift $v^{\mu}$ by an amount of order
$\Lambda_{\mathrm{QCD}}/m_Q$ as
\begin{equation}
p_Q^{\mu} =  m_Q v^{\mu} + k^{\mu} \rightarrow m_Q (v^{\mu}
+\frac{\epsilon^{\mu}}{m_Q} ) + k^{\mu}-\epsilon^{\mu} \equiv m_Q
v^{\prime \mu} +k^{\prime\mu}.    
\end{equation}
And the physics should be invariant under different decompositions of
the heavy quark momentum \cite{luke}.

In the collinear effective theory, the momentum of a collinear quark
is written as 
$P^{\mu} =\half(\overline{n} \cdot p) n^{\mu} +p_{\perp}^{\mu}
+k^{\mu}$. If we shift the largest quantity $\half(\overline{n}\cdot p)
n^{\mu}$ by a small amount of order $\lambda$ or $\lambda^2$, and if
this change is compensated by the change in $p_{\perp}^{\mu}
+k^{\mu}$, the total 
momentum remains unchanged and the physics should be invariant under
different decompositions of $P^{\mu}$. There is another ambiguity
because a different choice of the basis vectors $n^{\mu}$ and
$\overline{n}^{\mu}$ cannot change the physics as long as they satisfy 
$n^2=0$, $\overline{n}^2=0$ and $n\cdot \overline{n}=2$. However, a
small shift in momentum can be achieved by a small change in
$n^{\mu}$ and $\overline{n}^{\mu}$.

Manohar et al. \cite{manohar} have generalized the reparameterization
invariance, which was first considered in Ref.~\cite{chay1}, to three
classes of reparameterization transformations under which the basis
vectors $n^{\mu}$ and $\overline{n}^{\mu}$ change while $n^2=0$,
$\overline{n}^2=0$ and $n\cdot \overline{n}=2$ are fixed. The
effective Lagrangian in the collinear effective theory is invariant
under these transformations. The three classes of transformations are
given by
\begin{equation}
  \label{eq:genrep}
(\mathrm{I}) \left\{ \begin{array}{l}
n^{\mu}  \rightarrow n^{\mu} + \Delta^{\mu}_{\perp}, \\
\overline{n}^{\mu} \rightarrow \overline{n}^{\mu},
\end{array}
\right. \  (\mathrm{II}) \left\{ \begin{array}{l}
n^{\mu}  \rightarrow n^{\mu}, \\
\overline{n}^{\mu} \rightarrow
\overline{n}^{\mu}+\epsilon^{\mu}_{\perp}, 
\end{array}
\right. \  (\mathrm{III}) \left\{ \begin{array}{l}
n^{\mu}  \rightarrow (1+\alpha) n^{\mu}, \\
\overline{n}^{\mu} \rightarrow (1-\alpha) \overline{n}^{\mu}.
\end{array}
\right.
\end{equation}
Here $n\cdot \Delta_{\perp} = \overline{n}\cdot \Delta_{\perp} =
n\cdot \epsilon_{\perp} =\overline{n} \cdot \epsilon_{\perp} =0$ to
order $\lambda^2$. And the corresponding changes for arbitrary vectors
and quantum fields are summarized in Ref.~\cite{manohar}.

If we shift the largest component in the $n^{\mu}$ direction by a
magnitude of order $\lambda$ or $\lambda^2$, we can compensate this
shift by changing $p_{\perp}^{\mu} + k^{\mu}$ such that the total
momentum $P^{\mu}$ remains unchanged. The reparameterization
transformation is given by
\begin{equation}
  \label{eq:rep}
P^{\mu} = \frac{\overline{n} \cdot p}{2} \Bigl( n^{\mu}
+\frac{2\epsilon^{\mu}}{\overline{n}\cdot p} \Bigr)
+p_{\perp}^{\mu} +k^{\mu} -\epsilon^{\mu} \rightarrow
\frac{\overline{n} \cdot p}{2} 
n^{\prime 
  \mu} +p_{\perp}^{\prime \mu} +k^{\prime \mu} ,  
\end{equation}
with $\overline{n}^{\mu}$ fixed. The infinitesimal shift
$\epsilon^{\mu}$ includes momenta of order $\lambda$ or $\lambda^2$
with $n\cdot \epsilon =\overline{n}\cdot \epsilon=0$ to 
order $\lambda^2$. Here the last expression
shows that the shift in $n^{\mu}$ can be achieved using
the type-I transformation and $p^{\prime\mu}_{\perp}+k^{\prime\mu}=
p^{\mu}_{\perp} +k^{\mu} -\half (\overline{n}\cdot p)
\epsilon_{\perp}^{\mu} -\half (\epsilon_{\perp} \cdot p_{\perp})
\overline{n}^{\mu}$. The collinear quark field also changes as $\xi_n
\rightarrow \xi_n +\delta \xi_n$ to satisfy $\FMslash{n}
\xi_n=0$. Under the transformation
\begin{equation}
  \label{eq:reptr}
n^{\mu} \rightarrow n^{\mu} +\frac{2\epsilon^{\mu}}{\overline{n} \cdot
p}, \ \ \xi_n \rightarrow e^{i\epsilon  \cdot x} \Bigl( 1+
\frac{\FMslash{\epsilon}}{\overline{n} \cdot p}
\frac{\overline{\FMslash{n}}}{2} \Bigr) \xi_n,   
\end{equation}
the effective Lagrangian in Eq.~(\ref{eq:eff}) is invariant. Therefore
the collinear effective theory has a reparameterization invariance.
As a result, the kinetic energy term is not renormalized to all
orders in $\alpha_s$. 
For instance, the kinetic energy term to order $\lambda^2$, which is
given by 
\begin{equation}
  \label{eq:kin2}
K = \overline{\xi}_n \Bigl( n\cdot i\partial
+\frac{p_{\perp}^2}{\overline{n} \cdot p} + \frac{2p_{\perp} \cdot
  i\partial_{\perp}}{\overline{n}\cdot p}
+\frac{(i\partial_{\perp})^2}{\overline{n}\cdot p} -\frac{p_{\perp}^2
  i\overline{n} \cdot i\partial}{(\overline{n}\cdot p)^2} \Bigr)
\frac{\overline{\FMslash{n}}}{2} \xi_n,
\end{equation}
is not renormalized to all orders in $\alpha_s$. Here the first two
terms are of order $\lambda^0$, the third term is of 
order $\lambda$ and the last two terms are of order
$\lambda^2$.

The changes in $n^{\mu}$ of order $\lambda$ and $\lambda^2$ have
different implications on the structure of the effective theory
because the momentum has three scales in it. When we change $n^{\mu}$ by
an amount of order $\lambda$, it cannot be compensated by the residual
momentum $k^{\mu}$ which is of order $\lambda^2$, but can 
be compensated by the change in the label momentum
$p_{\perp}^{\mu}$. Therefore the change of order $\lambda$ induces a
reparameterization invariance between label momenta. This
reparameterization invariance is independent of the usoft fluctuation
of the fields. The reparameterization invariance at order $\lambda$
was utilized in deriving heavy-to-light current operators to order
$\lambda$ in Ref.~\cite{chay1}. 

If we change $n^{\mu}$ by an amount of order $\lambda^2$, then it can
be compensated by the residual momentum $k^{\mu}$. In terms of the
operator $\mathcal{P}^{\mu} = n^{\mu} \overline{\mathcal{P}}/2
+\mathcal{P}_{\perp}^{\mu}$, and $i\partial^{\mu}$, the transformation
in Eq.~(\ref{eq:reptr}) causes 
\begin{equation}
  \label{eq:lam2}
\mathcal{P}_{\mu} \rightarrow \mathcal{P}_{\mu} + \epsilon_{\mu}, \ \
i\partial_{\mu} \rightarrow i\partial_{\mu} -\epsilon_{\mu}.
\end{equation}
This implies that reparameterization invariant operators must be built
out of the linear combination $\mathcal{P}_{\mu} +i\partial_{\mu}$ and
the usoft gauge invariance requires that $i\partial_{\mu}$ be replaced
by the covariant derivative $iD^{\mu} =
i\partial^{\mu} -gA_u^{\mu}$.

There is another way to shift the large component $\half
(\overline{n}\cdot p) n^{\mu}$ by shifting $\overline{n}\cdot p$
instead of $n^{\mu}$. This
also has an analogue in HQET, namely, the residual mass effect
\cite{falk}. We can shift the large momentum by shifting the heavy
quark mass $m_Q$ as
\begin{equation}
  \label{eq:resmass}
p_Q^{\mu} = m_Q v^{\mu} + k^{\mu} \rightarrow (m_Q + \delta m) v^{\mu}
+ k^{\mu} - \delta m v^{\mu} \equiv m_Q^{\prime} v^{\mu} +
k^{\prime\mu},
\end{equation}
which is another legitimate decomposition into a large and a small
momenta as long as $\delta m$ is of order
$\Lambda_{\mathrm{QCD}}$. The effective theory using $m_Q^{\prime}$
should lead to the same results as the effective theory using
$m_Q$, and the combinations
$m_Q^* \equiv m_Q^{\prime} +\delta m, \ \ i\mathcal{D}^{\mu} \equiv
iD^{\mu} -\delta m v^{\mu}$  
are invariant under redefinitions of $\delta m$. As a consequence
hadronic form factors in the HQET must be defined in terms of matrix
elements containing the operator $i\mathcal{D}^{\mu}= iD^{\mu} -\delta
m\, v^{\mu}$. 
The physics should be unambiguous in defining a heavy quark mass and
the corresponding change in the HQET is called the residual mass
effect \cite{falk}.

In the collinear effective theory, we can change $\overline{n}\cdot p
\rightarrow \overline{n}\cdot p + 2\delta E$, which shifts
$\overline{n} \cdot p$ by an amount of $\lambda^2$. This
can be achieved by shifting $\overline{n}^{\mu}$ with fixed $n^{\mu}$,
which corresponds to the type-II transformation. We can decompose the
momentum of a collinear quark as 
\begin{eqnarray}
  \label{eq:cores}
P^{\mu} &=& \frac{\overline{n}\cdot p}{2} n^{\mu}  +p_{\perp}^{\mu} +
k^{\mu} \nonumber \\
&=& \Bigl(\frac{\overline{n}\cdot p}{2} -\delta E\Bigr) n^{\mu}
+p_{\perp}^{\mu} + k^{\mu}+\delta E n^{\mu} \rightarrow
\frac{\overline{n}^{\prime}\cdot p}{2} n^{\mu} 
+p_{\perp}^{\prime\mu} + k^{\prime\mu}, 
\end{eqnarray}
where the last expression shows that this change can be achieved by
the type-II transformation in Eq.~(\ref{eq:genrep}), along with the
corresponding change in $\xi_n$.
When we shift $\overline{n}_{\mu} \rightarrow \overline{n}_{\mu}
+\epsilon_{\mu}^{\perp}$, $\delta E$ is given by $\delta E = -
\epsilon_{\perp} \cdot p_{\perp}/2$, and
$p_{\perp}^{\prime\mu}+k^{\prime\mu} = 
p_{\perp}^{\mu} +k^{\mu} -\half (\epsilon_{\perp} \cdot p_{\perp})
n^{\mu}$. If $\epsilon_{\perp}$ is of order $\lambda$, $\delta E$ is
of order $\lambda^2$, which can be
compensated by the residual momentum $k^{\mu}$.
The final decomposition in Eq.~(\ref{eq:cores}) is also a legitimate
decomposition of the momentum $P^{\mu}$. And the
effective theory is invariant under this ``residual energy
transformation''. We call the invariance
under this transformation as ``residual energy invariance'' borrowing
the terminology from the HQET. 

Following the reasoning in HQET, the combinations
\begin{equation}
  \label{eq:cetinv}
\frac{\overline{n}\cdot p^*}{2} \equiv \frac{\overline{n}^{\prime}
  \cdot   p}{2} +\delta E, \ \ i\mathcal{D}^{\mu} \equiv iD^{\mu}
  -\delta E n^{\mu}  
\end{equation}
are invariant under redefinitions of $\delta E$. But here comes a
big difference between the HQET and the collinear effective
theory. In HQET, we have to use the combinations
$m_Q^* = m_Q^{\prime} +\delta m, \ \ i\mathcal{D}^{\mu} = iD^{\mu}
-\delta m v^{\mu}$,  
which are invariant under redefinitions of $\delta m$, and these
combinations should appear in the HQET Lagrangian. However,
the Lagrangian of the collinear effective theory is already invariant
under the residual energy transformation, which works even when  the
variation $\epsilon_{\mu}^{\perp}$ can be of order $\lambda^0$
\cite{manohar}. In other words, if we rewrite the Lagrangian in terms
of $\overline{n}\cdot p^*$ and $i\mathcal{D}^{\mu}$, there appears no
term depending on $\delta E$. Therefore it is irrelevant to use either
$\overline{n}\cdot p$, $iD^{\mu}$ or $\overline{n}\cdot p^*$,
$i\mathcal{D}^{\mu}$, and the physics is the same. This means that
there is no ambiguity in the 
choice of the large momentum component of order $\lambda^0$. And there
is no need to change the 
matrix elements containing the operator $iD^{\mu}$ and the
nonperturbative parameters in the calculation of form factors. This
need not be true for heavy-to-light currents.

We have considered various symmetries of the effective Lagrangian in
the collinear effective theory and their implications. Collinear and
usoft gluons are associated with collinear and usoft gauge
transformations. If we require the gauge invariance order by order in
$\lambda$, the gauge transformation of collinear gluons and
usoft gluons takes a definite form. The requirement of having the
gauge invariance at each order in $\lambda$ is important in
constructing gauge-invariant operators order by order. 

Reparameterization invariance imposes a constraint on the form of the
effective Lagrangian. The derivative term should appear in the
combination $\mathcal{P}^{\mu}+iD^{\mu}$, which relates quantities
with different powers of $\lambda$. Also the kinetic energy term is
not normalized to all orders in $\alpha_s$. Residual energy invariance
means that there is no ambiguity in defining the large
component. Though we change the largest momentum component by
order $\lambda$ or $\lambda^2$, the physical effect under this small
change does not appear.

We have considered heavy-to-light currents and their renormalization
behavior to order $\lambda$ in Ref.~\cite{chay1}. If we consider
heavy-to-light currents to order $\lambda^2$, reparameterization
invariance both in the collinear effective theory and the HQET will be
useful in constructing operators at order $\lambda^2$. If we use the
reparameterization transformation and the residual energy
transformation of order $\lambda$, they will change the heavy quark
velocity $v^{\mu} = (n^{\mu} +\overline{n}^{\mu})/2$ by an order
$\lambda$. This will confuse the power counting in $\lambda$ since the
energy of the 
light degrees of freedom in a heavy quark is of order
$\lambda^2$. However it is possible to make $v^{\mu}$ intact by
combining
the reparameterization transformation and the residual energy
transformation. For example, we consider both transformations at
order $\lambda^2$, or we can combine both transformations of order
$\lambda$ such that the changes
caused by these transformations cancel in such a way that the combination
$n^{\mu}+\overline{n}^{\mu}$ does not change. In this case, we can
combine the reparameterization invariance in the HQET and the
collinear effective theory to study the structure of
higher-dimensional heavy-to-light 
current operators. Studies in this direction are in progress.

{\bf Note added}: 

While this paper was being written,
Ref.~\cite{manohar} appeared. In Ref.~\cite{manohar}, the authors
extended the idea of reparameterization invariance to three possible
cases of changing the basis vectors $n^{\mu}$ and
$\overline{n}^{\mu}$. We note that the reparameterization
transformation and the residual energy transformation can be achieved
by using the type-I and the type-II transformations respectively. And
the type-III transformation is new.

\section*{Acknowledgments}
The authors are supported by Hacksim BK21 Project.

\end{document}